# ABSORPTION OF HIGH ENERGY GAMMA-RAYS BY LOW ENERGY INTERGALACTIC PHOTONS

F.W. STECKER
*Lab. for High Energy Astrophysics*
*NASA Goddard Space Flight Center*
*Greenbelt, MD 20771*

and

O.C. DE JAGER
*Dept. of Physics*
*Potchefstroom University*
*Potchefstroom 2520, South Africa*

**Abstract.** Following our previously proposed technique, we have used the recent $\gamma$-ray observations of Mkr421 to place theoretically significant constraints on and possible estimates of the intergalactic infrared radiation field (IIRF). Our $2\sigma$ upper limits and estimates are consistent with normal galactic IR production by stars and dust. They rule out exotic mechanisms proposed to produce a larger IIRF.
We find possible evidence for intergalactic absorption of $\gamma$-rays above 3 TeV energy in the observed spectrum of Mkr421. The implied IIRF is of the magnitude expected to be produced by stellar emission and reprocessing in galaxies. The sharpness of the spectral turnover is not consistent with the interpretation of this feature as a cutoff in the emission from the source. However, should the cutoff primarily be a result of absorption within the source itself, our possible extragalactic flux estimate becomes the strongest upper limit extant on the IIRF.
Using models for the low energy intergalactic photon spectrum from microwave to ultraviolet energies, we calculate the opacity of intergalactic space to $\gamma$-rays as a function of energy and redshift. These calculations indicate that the GeV $\gamma$-ray burst recently observed by the CGRO EGRET detector originates at a redshift less than $\sim 1.5$.

## 1. Introduction

The potential importance of the photon-photon pair production process in high energy astrophysics was first pointed out by Nikishov (1961). Unfortunately, his early paper overestimated the energy density of the low energy IIRF by about three orders of magnitude. However, with the discovery of the cosmic microwave background radiation, Gould and Schreder (1966) and Jelly (1966) were quick to point out the opacity of the universe to photons of energy greater than 100 TeV. Stecker (1969) and Fazio and Stecker (1970) generalized these calculations to high redshifts, showing that photons originating at a redshift $z$ will be absorbed above an energy of $\sim 100(1+z)^{-2}$ TeV. Following the detection of the blazar 3C279 at energies of 0.1 to 5 GeV by the EGRET team (Hartman, *et al.* 1992), we (Stecker, De Jager and Salamon 1992, Paper I) proposed that one look for pair-production absorption features in blazar spectra at TeV energies to determine the intensity of the IIRF. *Direct* observational determinations of the extragalactic infrared background radiation are fraught with difficulties (see section 3).

As we discussed in Paper I, an EGRET–strength grazar with a hard spectrum extending to multi-TeV energies is potentially detectable with ground-based telescopes. Pair-production interactions with the cosmic microwave background radiation will not cut off the $\gamma$-ray spectrum of a nearby (low redshift) source below an energy of $\sim 100$ TeV (Fazio and Stecker 1970). The techniques involved in making ground-based observations of TeV $\gamma$-ray sources have been reviewed by Weekes (1988).

Shortly after Paper I was published, the Whipple Observatory group reported the discovery of $\sim$ TeV $\gamma$-rays coming from the nearby blazar Mkr421, a BL Lac object (Punch *et al.* 1992). This source has a hard, roughly $E^{-2}$ $\gamma$-ray spectrum extending to the TeV energy range (Lin, *et al.* 1992; Mohanty, *et al.* 1993) where the pair production process involving infrared photons becomes relevant. Following the discovery of Mkr421, we made use of the TeV observations of this source to carry out our suggestion of using such sources to probe the IIRF (Stecker and De Jager 1993; De Jager, Stecker and Salamon 1994).

There are now over 40 grazars which have been detected by the EGRET team (Fichtel, *et al.* 1994). These sources, optically violent variable quasars and BL Lac objects, have been detected with a redshift range extending beyond 2. Of these objects, only the low reshift BL Lac, Mkr421 was seen by the Whipple telescope. The fact that the Whipple team did not detect the much brighter EGRET source 3C279 at TeV energies (Vacanti, *et al.* 1990, Kerrick, *et al.* 1993) is consistent with our previous predictions (Paper I; Stecker 1993) because this source is at a much higher redshift (z=0.54). However, the interpretation is somewhat complicated by the significant time-variability of this source at GeV energies and the fact that simultaneous measurements were not made.

The TeV spectrum of Mkr421 obtained by the Whipple Observatory group is consistent with a perfect $E^{-2.06}$ power-law, showing no absorption out to an energy of at least 3 TeV (Mohanty, *et al.* 1993). It is this lack of absorption below 3 TeV which we (Stecker and De Jager 1993) used to put upper limits on the extragalactic infrared energy density in the $\sim 0.2$ to 1 eV energy range ($\sim 1$ $\mu$m to 5 $\mu$m) (See also Dwek and Slavin 1994). However, there *is* possible evidence of absorption in the Mkr421 spectrum *above* $\sim 3$ TeV, which we have interpreted as possible intergalactic absorption allowing an estimate the IIRF at wavelengths longer than $\sim 5$ $\mu$m (De Jager, Stecker and Salamon 1994).

In this paper, we will review those results. We will also use semiempirical models of the low energy intergalactic radiation fields from microwave to ultraviolet energies to calculate the opacity of the universe as a function of redshift to $\gamma$-rays extending from the EGRET range (GeV energies) to energies in the PeV region. Using these calculations, we can place a limit on the redshift from which the multi-GeV $\gamma$-ray burst observed by EGRET can have originated.

## 2. The Absorption of TeV Photons by Infrared Photons

Consider the interaction of a TeV $\gamma$-ray of energy $E(z) = (1+z)E$ with a soft photon of energy $\epsilon(z) = (1+z)\epsilon$ at a redshift $z$, where $E$ and $\epsilon$ are the presently observed (z=0) photon energies. Pair production is expected above the threshold energy condition $E\epsilon(1+z)^2 x > 2(mc^2)^2$, where $x = (1-\cos\theta)$ and $\theta$ is the angle between the photon directions. The cross section is given by (Heitler 1960)

$$\sigma[E(z),\epsilon(z),x] = 1.25\times 10^{-25}(1-\beta^2)\left[2\beta(\beta^2-2)+(3-\beta^4)\ln\left(\frac{1+\beta}{1-\beta}\right)\right] \text{ cm}^2 \quad (1)$$

where $\beta = (1 - 2(mc^2)^2/(E\epsilon x(1+z)^2))^{1/2}$. For $\gamma$-rays in the TeV energy range, this cross section is maximized when the soft photon energy is in the infrared range:

$$\epsilon(E) \simeq \frac{2(mc^2)^2}{E} \simeq 0.5\left(\frac{1\text{ TeV}}{E}\right) \text{ eV} \quad (2)$$

For a $\sim 1$ TeV $\gamma$-ray, this corresponds to a soft photon wavelength near the K-band ($\sim 2\mu$m). Thus, infrared photons with wavelengths around $2\mu$m will contribute most to the absorption of TeV $\gamma$-rays. If the photons with energy $\epsilon$ have a number density $n(\epsilon)d\epsilon$ cm$^{-3}$, the optical depth for $\gamma$-rays will be

$$\tau(E) = \frac{c}{H_o}\int_0^z dz\,(1+z)(1+\Omega z)^{-1/2}\int_0^2 dx\frac{x}{2}\int_{\epsilon_1}^\infty d\epsilon\,n(\epsilon)\sigma\left[2x\epsilon E(1+z)^2\right] \quad (3)$$

where $\epsilon_1 = 2m^2c^4/Ex(1+z)^2$ (Stecker 1971).

We (Stecker and De Jager 1993) used the Whipple data on the TeV energy spectrum from Mkr421 to put limits on $\tau$, and therefore to put limits on the infrared radiation density. For a low redshift source ($z \ll 1$) such as Mk421, the calculation of the optical depth, $\tau$, simplifies considerably. Taking the soft photon spectrum as $n(\epsilon) = \kappa\epsilon^{-\alpha}$, for small redshifts, $z$, (with $\Omega = 1$)

$$\tau(E) = \frac{2^\alpha cz}{(\alpha+1)H_o}\left(\frac{2(mc^2)^2}{E}\right)n\left[\frac{2(mc^2)^2}{E}\right]\int_1^\infty u^{-\alpha}\,du\,\sigma[\beta = \sqrt{1-1/u}] \quad (4)$$

For normal galaxies (Stecker et al. 1977) with $n(\epsilon) = \kappa_1\epsilon^{-2.55}$, in units of eV$^{-1}$cm$^{-3}$ with $\epsilon$ in eV, we find

$$\tau(E) = 143\kappa_1 E_{\text{TeV}}^{1.55} \quad (5)$$

and, for $\epsilon u_\epsilon$ as constant (or $n(\epsilon) = \kappa_2 \epsilon^{-2}$), we find that

$$\tau(E) = 120\kappa_2 E_{\text{TeV}}. \tag{6}$$

Expressions (5) and (6) for $\tau(E)$ should roughly hold up to $E \sim 10$ TeV.

Using the 11 spectral points then available, Mohanty et al. (1993) were able to fit a single power law between their data set below 5.1 TeV and the EGRET data set at $\sim$ GeV energies (Lin, et al. 1992). The spectrum that they obtained was

$$dN/dE = (1.02 \pm 0.14) \times 10^{-11} E^{-2.06 \pm 0.04} \text{ photons/cm}^2/\text{s}/\text{TeV} \tag{7}$$

We generalized this fit to include absorption by assuming spectra of the form

$$dN/dE = K E^{-\Gamma} \exp\left[-\tau(E)\right] \tag{8}$$

with $K$, $\Gamma$ and $\kappa$ as free parameters. Then the best $\chi^2$ fit is for

$$dN/dE = (1.11 \pm 0.21) \times 10^{-11} E_{\text{TeV}}^{-2.04 \pm 0.04} \exp(-\tau(E)) \tag{9}$$

where (assuming $H_o = 75$ km/s/Mpc) $\tau(E) = 0.048 E_{\text{TeV}}$ (for $\alpha = 2$) and $\tau(E) = 0.043 E_{\text{TeV}}^{1.55}$ (for $\alpha = 2.55$) (again valid only up to $E \sim 10$ TeV) with a minimum $\chi_8^2 = 4.9$ for $11-3 = 8$ degrees of freedom for both models for $n(\epsilon)$. The upper limits on $\tau$ are obtained by allowing all free parameters to vary simultaneously, which allows us to construct confidence ellipsoids. The $1\sigma$ and $2\sigma$ upper limits on $\kappa_1$ in Eq. (5) (i.e. for $\alpha = 2.55$) are $\kappa_1 < 9 \times 10^{-4}$ and $\kappa_1 < 1.9 \times 10^{-3}$ respectively, whereas the corresponding numbers for $\kappa_2$ in Eq. (6) (i.e. for $\alpha = 2$) are $\kappa_2 < 2.1 \times 10^{-3}$ and $\kappa_2 < 4.0 \times 10^{-3}$ respectively. The $2\sigma$ or 95% upper limits on $\tau(E)$ are then $\tau(E) < 0.27 E_{\text{TeV}}^{1.55}$, which implies $n(\epsilon) < 1.9 \times 10^{-3} \epsilon^{-2.55}$ eV$^{-1}$cm$^{-3}$ for $\alpha = 2.55$. For $\alpha = 2$ we find $\tau(E) < 0.48 E_{\text{TeV}}$, which implies $n(\epsilon) < 4.0 \times 10^{-3} \epsilon^{-2}$ eV$^{-1}$cm$^{-3}$. The latter upper limit (for $H_o = 50$ km/s/Mpc) is shown on Fig. 1.

In the above analysis, the spectral information *above* 3.4 TeV was neglected since the highest energy points were only upper limits. However, we note that using the data from Mohanty et al. (1993) and extrapolating the $E^{-2.06}$ differential spectrum to energies above 3 TeV, we expect $\sim 58$ ON-source events (source $\gamma$-rays and cosmic-ray background) between 3.4 TeV and 5.1 TeV, whereas only 26 *on*-source events were observed, which is consistent with the 27 *off*-source cosmic ray background events collected with similar exposure. Furthermore, above 5.1 TeV we expect $\sim 32$ *on*-source events, whereas only 10 were observed against the 12 background events. These data may be interpreted as indicating the presence of an absorption feature in the *integral* spectrum of Mkr421 manifested at the highest energies. Such an interpretation allows us to obtain an estimate of the IIRF, rather than an upper limit (De Jager, Stecker and Salamon 1994). This empirical estimate is consistent with theoretical estimates of the IIRF (Stecker, *et al.* 1977).

Even though the EGRET and Whipple detections were not simultaneous, they can be cautiously combined. Although the intensity of Mkr421 may occasionaly vary by large amounts on a timescale of days (Kerrick, *et al.* 1995), it was shown by the Whipple group that averaging on a timescale of several weeks results in a 20% variation at most. Such a timescale is the one relevant to the above results.

There are three main caveats to consider with regard to the interpretation by De Jager, *et al.* (1994) that the rollover in the high end of the observed spectrum of Mkr421 is an absorption effect of the IIRF: (1) There may be some uncertainty in determining the event rate derived by Mohanty, *et al.* (1993) in the energy range above 5 TeV, owing to a possible *oversubtraction* of assumed cosmic ray produced events, which can become more confused with $\gamma$-ray produced events at the higher energies (R. Lamb, priv. comm.). We note, however, that the Crab nebula, a much stronger source, was observed by the Whipple group to have a power–law spectrum with *no* such turnover up to $\sim$ 10 TeV (Vacanti, *et al.* 1991). (2) The source spectrum may have an *intrinsic emission* cutoff above $\sim$ 3 TeV. However, one may then ask why the intrinsic spectrum would turn over at the energy naturally expected for an IIRF absorption feature. We also note that emission models involving Compton scattering of high energy electrons in the source would not produce such a sharp cutoff (Ghisellini, Maraschi and Treves 1985; Dermer and Schlickheiser 1993; Bloom and Marscher 1993). (3) *Intrinsic absorption* in the source may produce a high–energy cutoff. However, we consider it highly improbable that the infrared characteristics of the source would conspire to exactly mimic that which would be produced by a reasonably estimated IIRF. Also, if the observed IR flux from Mkr421 (Makino, *et al.* 1987) is from the galaxy, the implied optical depth would not be large enough to produce an observable absorption effect, while if it originates in the jet and is beamed, the intrinsic flux may be considerably reduced as well as the interaction rate. We note, however, that if the absorption *is* intrinsic, then our estimate becomes the best upper limit extant on the IIRF.

We also note that intergalactic electron-photon cascading (Protheroe and Stanev 1993; Aharonian, Coppi and Volk 1993) is not expected to significantly influence the TeV spectrum of Mkr421 since the source is observed within a point-spread function with HWHM of $\sim$ 0.2 degrees, and it can be shown that an intergalactic magnetic field of $10^{-20}$ G or greater would deflect the cascade TeV electrons away from the source direction, creating a $\gamma$-ray halo much larger than the point-spread function. Since the measured magnetic field strengths in superclusters are of the order of a $\mu$G (Kronberg 1994 and references therein), it is highly unlikely that the intergalactic field would be as weak as $10^{-20}$ G.

Thus, we conclude our interpretation of the Mkr 421 spectrum may indeed be the first determination of the IIRF at a level which is expected from stellar processes in galaxies (Stecker, *et al.* 1977).

## 3. The Intergalactic Infrared Background Radiation Field

A theoretical upper limit for the infrared energy density from starlight in normal galaxies was estimated by Stecker, *et al.* (1977), assuming that all of the energy from nucleosynthesis in stars is emitted in the infrared. Estimates of the intergalactic and optical radiation densities were obtained by Yoshii and Takahara (1988), Takahara (1990), and Tyson (1990), based on galaxy counts and other modelling considerations. Cowie, *et al.* (1990) have estimated the IIRF using a K-band deep galaxy survey.

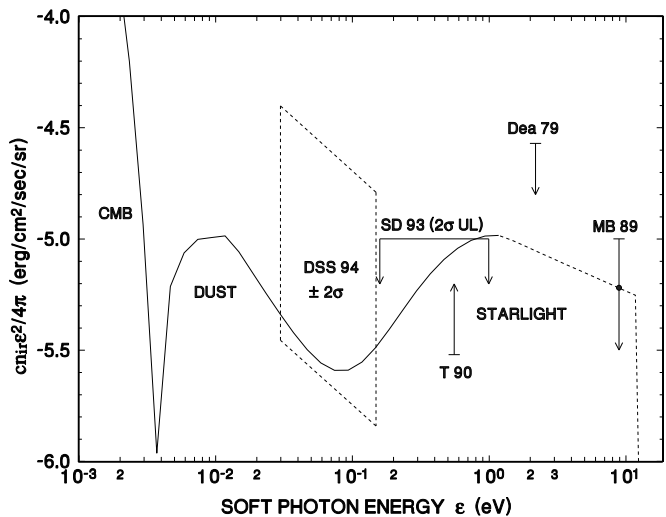

Fig. 1. The low energy intergalactic photon spectrum from the microwave to the Lyman limit in the ultraviolet modeled by including the cosmic microwave background (CMB) spectrum and dust and starlight spectra modeled by Franceschini, *et al.* (1994) extrapolated into the UV using the results of Martin and Bowyer (1989), indicated as 'MB 89'. The lower limit given by Tyson (1990) is indicated as 'T 90' and the upper limit of Dube, *et al.* (1979) is indicated as 'Dea 79'. We also show our previous upper limit (Stecker and De Jager 1993, 'SD 93') and our estimate of the IIRF (De Jager, Stecker and Salamon 1994, 'DSS 94'), obtained from the TeV data on Mkr421 (see text).

Direct measurements of the extragalactic infrared and optical radiation are plagued by the difficulty of separating the true extragalactic component from galactic radiation, zodiacal light and instrumental background (e.g.

radiation from rocket exhaust). The observational upper limits obtained in the near infrared by Matsumoto, *et al.* (1988) and the optical (Dube, *et al.* 1979, see Fig. 1) all lie well above the theoretical starlight spectrum shown in Fig. 1, as well as the other estimates mentioned above. Thus, it is clear that the present directly obtained upper limits are not theoretically significant. However, as we discussed in the previous section, already the TeV data from Mkr421 have provided both estimates and upper limits which *are* theoretically significant. In fact, the estimates of De Jager, Stecker and Salamon (1994) indicate that at least 30% of the energy emitted by stars and dust in galaxies winds up in the infrared.

The intergalactic background radiation from galactic starlight and dust reradiation has recently been modeled in some spectral detail (Franceschini, *et al.* 1994; MacMinn and Primack 1994). As discussed by Franceschini, *et al.* (1994), our results are consistent with their theoretical models. Using their models for stellar and dust emission, together with the COBE results on the cosmic microwave-FIR background radiation and extrapolating to the Lyman limit in the UV, we obtain a full spectrum low energy intergalactic photon background as shown in Fig. 1. To obtain the UV estimate shown in Fig. 1, we have used the results of Martin and Bowyer (1989). It should be kept in mind that this UV estimate is very uncertain and represents an upper limit on the UV flux. This flux may actually drop off significantly at energies below the Lyman limit. (To allow for this possibility in our calculations below, we have used two other model estimates.)

Fig. 1 also shows our previously obtained upper limit (Stecker and De Jager 1993) and IIRF estimate (De Jager, Stecker and Salamon 1994). It can be seen that our estimate from the Mkr421 data lies somewhat above the model. This can be from either of three causes: (1) uncertainty in the interpretation of the Mkr421 event rate (R. Lamb, private communication), (2) some intrinsic absorption in Mkr421 (see previous discussion), or an underestimate in the modeled IIRF owing to the omission of contributions from infrared bright extragalactic sources (De Zotti, *et al.* 1994).

## 4. The Opacity of the Universe to Gamma-Rays

By making use of the modeled intergalactic spectrum shown in Fig. 1, one can calculate the opacity of the universe at various redshifts over a wide range of $\gamma$-ray energies. Fig. 2 shows the optical depth of the universe to $\gamma$-rays originating at the redshift corresponding to the grazars designated. We can also use the low energy intergalactic photon spectrum shown in Fig. 1 to obtain a redshift limit on the 17 February 1994 high energy $\gamma$-ray burst observed by EGRET (Hurley, *et al.* 1994). The highest energy photon in this burst had an energy of $18 \pm 5$ GeV. The corresponding opacity to photons in this energy range is shown in Fig. 3. In all of the models of the intergalactic UV background which we used (see figure caption), the universe becomes opaque to $\sim 20$ GeV photons at a redshift of $\sim 1.5$. In all of these models, we assumed that there are no intergalactic ionizing photons above the Lyman limit of 13.6 eV. The probable existence of such photons would only decrease our estimate of the absorption redshift. Therefore, one obtains the interesting result that the source of this burst lies at a redshift which is not uncommon for high energy astrophysical sources. One should be cautioned, however, that the existence of one burst in this redshift range does not necessarily rule out exotic cosmological models for $\gamma$-ray bursts, such as those involving topological defects; a string located at a "low" redshift would be brighter and more observable at high energies than one at very high redshifts.

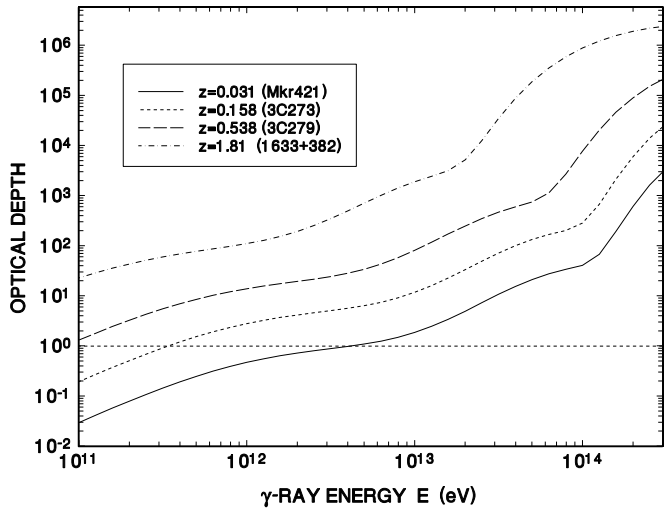

Fig. 2. Optical depth versus energy for photons from four EGRET grazars obtained using the photon spectrum shown in Fig. 1 and taking $H_o = 50$ km/s/Mpc.

We have used the opacity curve shown in Fig. 2 and $H_o = 50$ km/s/Mpc to fit the Whipple data on Mkr421 as shown in Fig. 4. The Whipple spectrum is well fitted by a simple $E^{-2}$ source spectrum with absorption included ($\chi^2 = 7.7$ for $8 - 2 = 6$ degrees of freedom). Fig. 4 also shows the latest totaled EGRET spectra (Lin, et al. 1994). It can be seen that an extrapolation of our Mkr421 fit on the TeV data will adequately agree with the EGRET results.

Again, using the opacity curves shown in Fig. 2, we have calculated predicted turnovers in other EGRET observed grazars at various redshifts, assuming power-law spectra with the indices fitted by the EGRET team (Fichtel, et al. 1994). Our results are shown in Fig. 5. It can be seen that we predict turnovers in such spectra which should be observable with sensitive enough multi-GeV and sub-TeV detectors and used in concert to test our model of intergalactic absorption.

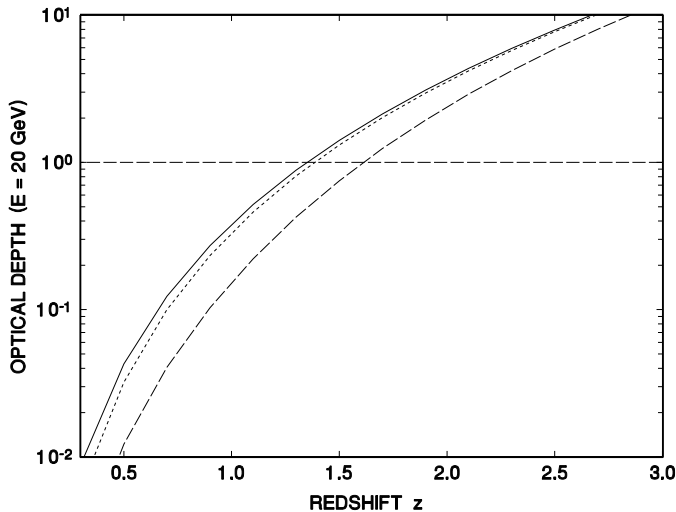

Fig. 3. Optical depth versus redshift for photons of energy 20 GeV (assuming $H_o = 50$ km/s/Mpc). The *solid line* was obtained using the IIRF model shown in Fig. 1. The *short dashed line* was obtained by modifying the UV model with an exponential cutoff of the form $\exp(-\epsilon/7.7 \text{ eV})$ above 7.7 eV. For the *long dashed line* the starlight component between 0.1 eV and 1.2 eV was mirrored into the range 1.2 to 13.6 eV, resulting in symmetric peak around 1.2 eV. In all cases, we took an absolute cutoff in the UV at the Lyman limit of 13.6 eV. The results shown can be used to put an upper limit on the redshift of the 17 February 1994 burst observed by EGRET (see text).

We have also pointed out that by combining a measurement of the opacity of the universe to high energy $\gamma$-rays with a direct measurement of the IIRF, one can determine the value of the Hubble parameter $H_0$ at cosmologically significant distances, independent of the distance ladder (Salamon, Stecker and De Jager 1994). Such a determination of $H_0$ has become even more critical with the recent, relatively "local" measurement of $H_0$ using Cepheids in the Virgo cluster (Pierce, *et al.* 1994; Freedman, *et al.* 1994). These determinations using observations form the Hubble Space Telescope give values of $H_0 \sim 80-90$ km s$^{-1}$ Mpc$^{-1}$, resulting in a *prima facie* conflict with age determinations of old stars in globular clusters. A determination of a smaller value for $H_0$ sampled at truly cosmological redshifts could resolve this conflict. Indeed, it has very recently been suggested by Wu, *et al.* (1994) that the "local" value of $H_0$ could be 1.2 to 1.4 times larger than its true cosmological value, owing to gravitational effects.

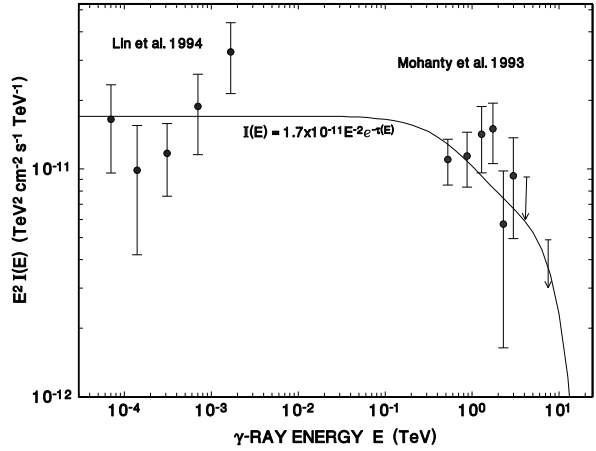

Fig. 4. Mkr421 spectrum fitted to the Whipple data (Mohanty, et al. 1993) with absorption calculated using the model spectrum in Fig. 1 and $H_o = 50$ km/s/Mpc. The upper limits above 3 TeV are at the $2\sigma$ level. The last point follows from the integral flux above 5 TeV, which was converted to a differential point given the expected absorption spectrum. Also shown are the most recent EGRET data points (Lin, et al. 1994).

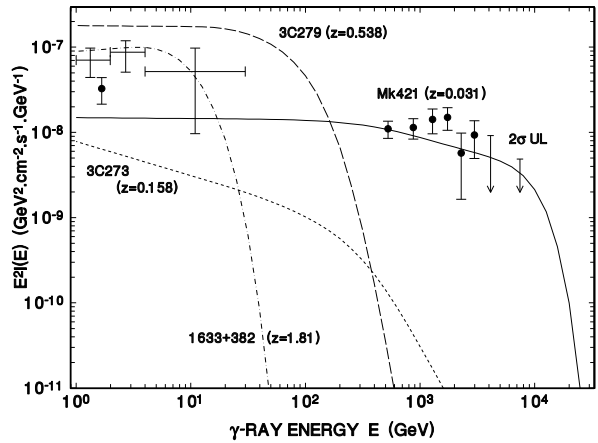

Fig. 5. Power-law spectra extapolated from EGRET data for four observed grazars. The absorption calculations assumed $H_o = 50$ km/s/Mpc. The solid circles are data points for Mkr421. The plain crosses are data points for the high redshift grazar 1633+382 (Mattox, et al. 1993).

## 5. Conclusions

The observational upper limits on the extragalactic infrared background radiation field are not very useful in constraining the various theoretical models proposed. We (Stecker, et al. 1992) have suggested that using sub-TeV and multi-TeV observations of grazars, one can derive useful information about the extragalactic infrared background provided that the intrinsic source spectrum does not cut off in this energy range. This appears to be the case with Mkr421. If the spectrum of an extragalactic source extends beyond 1 TeV, then the extragalactic infrared field should cut off the observed spectrum between $\sim 10$ GeV and a few TeV, depending on the redshift of the source. There is possible evidence for such a sharp drop in the spectrum of Mkr421 above 3 TeV where intergalactic absorption is expected to occur. We have used this evidence to estimate the intensity of the IIRF.

Our conservative upper limits on the IIRF, obtained from the Whipple data on Mkr421 below 3 TeV, rule out various exotic mechanisms for producing larger fluxes, such as some exploding star, decaying particle, massive object and black hole models (Carr 1988). They are consistent with the extragalactic near infrared background originating from ordinary stellar processes in galaxies (Stecker, *et al.* 1977; Franceschini *et al.* 1994).

The background flux in the 1 to 5 $\mu m$ range both estimated and theoretically expected to be in the range $\sim 10^{-9}$ to $\sim 10^{-8}$ w m$^{-2}$ sr$^{-1}$. Such a flux may be eventually detectable with the DIRBE experiment on COBE, provided that the foreground radiation from zodiacal light can be modeled well enough to be subtracted out (Mather, 1982; Hauser 1990; Mather, *et al.* 1990). If this flux can be directly detected, it may be used in combination with high energy $\gamma$-ray absorption measurements to determine the value of the Hubble parameter at cosmologically significant redshifts (Salamon, Stecker and De Jager 1994). Our estimates of the opacity of the universe to $\gamma$-rays can also be used to constrain the redshifts of high energy $\gamma$-ray burst sources. The highest energy burst observed by EGRET is constrained to originate at a redshift less than $\sim 1.5$.

It would be desirable to extend $\gamma$-ray observations to sub-TeV energies and to grazars at differing redshifts with higher sensitivity instruments in order to accurately determine the low energy intergalactic background radiation fields (see Fig. 5).

## Acknowledgements

We thank M.H. Salamon for discussions of the manuscript.